\documentclass[twocolumn,prb,showpacs,superscriptaddress,groupedaddress]{revtex4}
\usepackage{epsfig,color}
\usepackage{amssymb}
\usepackage{dcolumn}
\usepackage{amstext}
\usepackage{graphicx}

\def\ra{\rangle}
\def\la{\langle}

\def\be{\begin{equation}}
\def\ee{\end{equation}}
\def\ba{\begin{eqnarray}}
\def\ea{\end{eqnarray}}

\begin{document}

\title{Disorder induced transition between $s_{\pm}$ and $s_{++}$ states in
two-band superconductors}
\author{D.V.~Efremov}
\email{efremov@fkf.mpg.de}
\affiliation{Max-Planck-Institut f\"{u}r Festk\"{o}rperforschung, D-70569 Stuttgart,
Germany}
\author{M.M.~Korshunov}
\email{korshunov@phys.ufl.edu}
\affiliation{Department of Physics, University of Florida, Gainesville, Florida 32611, USA}
\affiliation{L. V. Kirensky Institute of Physics, Siberian Branch of Russian Academy of Sciences, 660036 Krasnoyarsk, Russia}
\affiliation{Siberian Federal University, Svobodny Prospect 79, 660041 Krasnoyarsk, Russia}
\author{O.V.~Dolgov}
\affiliation{Max-Planck-Institut f\"{u}r Festk\"{o}rperforschung, D-70569 Stuttgart, Germany}
\author{A.A.~Golubov}
\affiliation{Faculty of Science and Technology and MESA+ Institute of Nanotechnology, University of Twente, 7500 AE Enschede, The Netherlands}
\author{P.J.~Hirschfeld}
\affiliation{Department of Physics, University of Florida, Gainesville, Florida 32611, USA}
\date{\today}

\begin{abstract}
We have reexamined the problem of disorder in two-band
superconductors, and shown within the framework of the  $T$-matrix
approximation, that the suppression of $T_c$ can be described by a
single parameter depending on the intraband and interband impurity
scattering rates. $T_c$ is shown to be more robust against
nonmagnetic impurities than would be predicted in the trivial
extension of Abrikosov-Gor'kov theory. We find a
disorder-induced transition from the $s_{\pm}$ state to a gapless
and then to a fully gapped $s_{++}$ state, controlled by  a single
parameter -- the  sign of the average coupling constant $\langle
\lambda \rangle$. We argue that this transition has strong
implications for experiments.
\end{abstract}

\maketitle

\textit{Introduction.} The symmetry and structure of the
superconducting order parameter in recently discovered iron-based
superconductors (FeSC) is one of the main challenges in this
 field~\cite{rev}. The Fermi surface (FS) is usually given
by two small hole pockets around the $\Gamma=(0,0)$ point and two
electron pockets around the $M=(\pi,\pi)$ point in the 2-Fe Brillouin zone. The
proximity of the competing spin-density-wave (SDW) state with $\mathbf{Q}=(\pi,\pi)$
suggests antiferromagnetic fluctuations as a mechanism for
electron pairing. In this
case, the natural order parameter for most of the FeSC is the so-called $%
s_{\pm}$ state, described by an isotropic order parameter on each FS with
the opposite signs for electronlike and holelike pockets. Many experimental
results, such as the NMR spin-lattice relaxation rate, the spin-resonance peak at
the SDW wave vector $\mathbf{Q}$ in inelastic neutron scattering, and
quasiparticle interference in tunneling experiments, are in good qualitative
agrement with this scenario, although some materials are more anisotropic
than others~\cite{reviews}.

Since varying amounts of disorder are present in the materials, and because
superconductivity is created in some cases by doping, it is important to
understand the role of impurities. It has been shown that in an $s_\pm$
state, any nonmagnetic impurity which scatters \textit{solely} between the
bands with a different sign of the order parameter suppresses $T_c$ in the
same way as a magnetic impurity in a single-band BCS superconductor~\cite{golubov97}. Therefore the critical temperature $T_c$ should obey the Abrikosov-Gor'kov (AG) formula $\ln T_{c0}/T_{c} = \Psi(1/2+ \Gamma/2\pi T_c) - \Psi(1/2)$, where $\Psi(x)$ is the digamma function and $T_{c0}$ is the critical temperature in the absence of impurities~\cite{AG}. The critical value of the scattering rate $\Gamma$ defined by $T_c(\Gamma^\mathrm{crit}) = 0$ is given by $\Gamma^\mathrm{crit}/T_{c0}= \pi/2 \gamma \approx 1.12$ within AG
theory. However in several experiments on FeSC, e.g. Zn substitution or
proton irradiation, it is found~\cite{cheng,li,nakajima,tropeano} that the $T_c$ suppression is much less than expected in the framework of AG theory.
It has therefore been suggested that the $s_\pm$ state is not realized at
all in these systems, and that a more conventional two-band order parameter
without sign change ($s_{++}$) is the more likely ground state~\cite{kontani,bang}.

The disorder problem in these systems is substantially more complicated than
this simple argument suggests, however. Even within the assumption of
isotropic gaps on two different Fermi pockets and nonmagnetic scattering, a
much slower pair-breaking rate can be achieved by assuming that the
scattering is primarily intraband rather than interband. In the pure
intraband scattering limit, Anderson's theorem applies, the system is
insensitive to the sign of the order parameter, and no $T_c$ suppression
occurs. The rate of $T_c$ suppression therefore apparently depends on the
interplay of both intraband and interband scattering rates, and drawing
conclusions regarding the superconducting state based on systematic disorder
studies is fundamentally more difficult than in one-band systems. One
approach to this problem has been to try to determine the intraband and
interband scattering potentials microscopically for each type of impurity
and host~\cite{KemperCo,AritaIkeda,mazin}, but the quantitative
applicability of band theory to such questions is unclear.

Here we consider the critical temperature of an isotropic $%
s_\pm$ two-band superconductor within the usual self-consistent
$T$-matrix approximation for impurity scattering~\cite{allen}. We
perform the study both analytically in the weak-coupling regime
and numerically in the strong-coupling Eliashberg framework. We
find that the dependence of $T_c$ on impurity concentration is
given by a universal form independent of impurity potentials, with
respect to a generalized pair-breaking parameter. The form depends,
however, on the ratio of interband to intraband pairing matrix
elements. Depending on the average values of these matrix
elements, we find there are two possible types of $s_\pm$
superconductivity. The first is the one which has been largely
discussed so far in the literature, for which $T_c $ is suppressed
as disorder is increased, until it vanishes at a critical value of
the scattering rate. There is however also a second type of
$s_\pm$ state, one for which $T_c$ tends to a finite value as
disorder increases; at the same time the gap function acquires a
uniform sign, i.e., undergoes a transition from $s_\pm$ to
$s_{++}$.

\textit{Model.} We consider the Eliashberg equations~\cite{allen} for a
two-band superconductor with a $4 \times 4$ matrix quasiclassical Green's
function in Nambu and band space,
\begin{equation}
\hat\mathbf{g}(\omega_n)=\left(
\begin{array}{cc}
g_{0a} & 0 \\
0 & g_{0b}%
\end{array}%
\right) \otimes \hat\tau_0 + \left(
\begin{array}{cc}
g_{1a} & 0 \\
0 & g_{1b}%
\end{array}
\right)\otimes \hat\tau_2,  \label{eq.g}
\end{equation}
where the $\tau_i$ denote Pauli matrices in Nambu space, and $g_{0\alpha}$
and $g_{1\alpha}$ are the normal and anomalous $\xi$-integrated Nambu Green's
functions:
\begin{equation}
g_{0\alpha} = - \frac{{\mathrm{i}} \pi N_\alpha \tilde\omega_{\alpha n}} {%
\sqrt{\tilde\omega_{\alpha n}^2 + \tilde\phi_{\alpha n}^2}}, \;\;
g_{1\alpha} = - \frac{ \pi N_\alpha \tilde\phi_{\alpha n}} {\sqrt{%
\tilde\omega_{\alpha n}^2 + \tilde\phi_{\alpha n}^2}}.  \label{g}
\end{equation}
Here, index $\alpha$ runs over band indices $a$ and $b$, $N_{a,b}$ are the
density of states of each band ($a$, $b$) at the Fermi level, and $\omega_n =
\pi T (2n+1)$ is the Matsubara frequency. The quantities $%
\tilde\omega_{\alpha n}$ and $\tilde\phi_{\alpha n}$ are Matsubara
frequencies and order parameters renormalized by the self-energy $\hat%
\mathbf{\Sigma}({\mathrm{i}}\omega_n)$, respectively,
\begin{eqnarray}
\tilde\omega_{\alpha n} &=& \omega_n + {\mathrm{i}} \Sigma_{0\alpha}({%
\mathrm{i}}\omega_n) + {\mathrm{i}} \Sigma_{0\alpha}^{imp}({\mathrm{i}}%
\omega_n),  \label{eq.omega.tilde} \\
\tilde\phi_{\alpha n} &=& \Sigma_{1\alpha}({\mathrm{i}}\omega_n) +
\Sigma_{1\alpha}^{imp}({\mathrm{i}}\omega_n).  \label{eq.Delta.tilde}
\end{eqnarray}
The self-energy due to the spin fluctuation interaction is then given by:
\begin{eqnarray}
\Sigma_{0\alpha}({\mathrm{i}}\omega_n) =  T
\sum\limits_{\omega_n^{\prime },\beta} |\lambda_{\alpha\beta}(n-n^{\prime
})| g_{0\beta}/N_\beta,  \label{eq.DeltaN2} \\
\Sigma_{1\alpha}({\mathrm{i}}\omega_n) = -T \sum\limits_{\omega_n^{\prime
},\beta} \lambda_{\alpha\beta}(n-n^{\prime }) g_{1\beta}/N_\beta.
\label{eq.DeltaN1}
\end{eqnarray}
The coupling functions $\lambda_{\alpha\beta}(n-n^{\prime }) =
2\lambda_{\alpha\beta} \int^{\infty}_{0} d\Omega \Omega B(\Omega)/\left[%
(\omega_n-\omega_{n^{\prime }})^{2}+\Omega^{2}\right]$ are expressed via the
spectral functions $B(\Omega)$ (Ref. \cite{parker}) and constants $\lambda_{\alpha
\beta}$. The matrix elements $\lambda_{\alpha \beta}$ can be positive (attractive) as
well as negative (repulsive) due to the interplay between spin fluctuations and
electron-phonon coupling~\cite{BS,parker} and strongly renormalized due to
the nested Coulomb interaction~\cite{chubukov}.

We use the $T$-matrix approximation to calculate the average impurity
self-energy  $\hat\mathbf{\Sigma}^{imp}$:
\begin{equation}
\hat\mathbf{\Sigma}^{imp}({\mathrm{i}}\omega_n) =n_{imp} \hat\mathbf{U} + \hat\mathbf{U} \hat%
\mathbf{g}(\omega_n)
\hat\mathbf{\Sigma}^{imp}({\mathrm{i}}\omega_n),
\label{eq.tmatrix}
\end{equation}
where $\hat\mathbf{U} = \mathbf{U} \otimes \hat\tau_3$ and $n_{imp}$ is
impurity concentration. For simplicity intraband and interband parts of the
potential are set equal to $v$ and $u$, respectively, such that $(\mathbf{U}%
)_{\alpha \beta} = (v-u) \delta_{\alpha \beta} + u$. This completes the
specification of the equations which determine the quasiclassical Green's
functions.

Note that we have neglected possible anisotropy in each order parameter $%
\tilde\phi_{a(b)n}$; these effects can lead to changes in the response of
the two-band $s_\pm$ system to disorder and have been examined, e.g. in Ref.~%
\onlinecite{Mishra}.

\textit{Critical temperature.} $T_{c}$ is found by solving the linearized
Eliashberg equations for the renormalization factors $\tilde{Z}_{\alpha n}=%
\tilde{\omega}_{\alpha n}/\omega _{n}$ and gap functions $\tilde{\Delta}%
_{\alpha n}=\tilde{\phi}_{an}/\tilde{Z}_{\alpha n}$~\cite{allen}:
\begin{eqnarray}
\tilde{Z}_{\alpha n} &=&1+\sum_{\beta }\tilde{\Gamma}_{\alpha \beta
}/|\omega _{n}|  \nonumber \\
&+&\pi T_c \sum\limits_{\omega _{n^{\prime }},\beta }|\lambda _{\alpha \beta
}(n-n^{\prime })|\mathrm{sgn}\left( {\omega _{n^{\prime }}}\right) /\omega
_{n},  \label{eq:Elias1} \\
\widetilde{Z}_{\alpha n}\tilde{\Delta}_{\alpha n} &=&\sum_{\beta }\tilde{%
\Gamma}_{\alpha \beta }\tilde{\Delta}_{\beta n}/\left\vert \omega
_{n}\right\vert   \nonumber \\
&+&\pi T_c \sum\limits_{\omega _{n^{\prime }},\beta }\lambda _{\alpha \beta
}(n-n^{\prime })\tilde{\Delta}_{\beta n^{\prime }}/|\omega _{n^{\prime }}|,
\label{eq:Elias2}
\end{eqnarray}%
where $\tilde{\Gamma}_{\alpha \beta }$ are impurity scattering rates.

If one inserts Eq.~(\ref{eq:Elias1}) into Eq.~(\ref{eq:Elias2}) and gets a set of equations for $\tilde{\Delta}_{\alpha n}$,
it is easy to show that the impurity intraband scattering terms $\propto \tilde{\Gamma}_{aa}$ and $\tilde{\Gamma}_{bb}$ drop out~\cite{EPAPS},
in agreement with Anderson's theorem.
From Eq.~(\ref{eq.tmatrix}) one finds $\tilde{\Gamma}_{ab(ba)}$ as
\begin{equation}
\tilde{\Gamma}_{ab(ba)}=\Gamma _{a(b)}\frac{(1-\tilde{\sigma})}{\tilde{\sigma%
}(1-\tilde{\sigma})\eta \frac{(N_{a}+N_{b})^{2}}{N_{a}N_{b}}+(\tilde{\sigma}%
\eta -1)^{2}},  \label{eq:f}
\end{equation}%
where $\tilde{\sigma}=(\pi ^{2}N_{a}N_{b}u^{2})/(1+\pi ^{2}N_{a}N_{b}u^{2})$
and $\Gamma _{a(b)}=%2
n_{imp}\pi N_{b(a)}u^{2}(1-\tilde{\sigma})$ are generalized cross-section
and normal state scattering rate parameters, respectively. The parameter $\eta $ controls the ratio of intra-band and inter-band scattering as
 $v^2=u^2\eta $.
In the {\it Born }(weak scattering) {\it limit}, $\tilde{\sigma}\rightarrow 0$, while for $\tilde{\sigma}\rightarrow 1$ the {\it unitary limit} (strong
scattering) is achieved. From (\ref{eq:f}), we therefore recover explicitly
the well-known but counterintuitive result that in the unitary limit
nonmagnetic impurities do not affect $T_{c}$ in an $s_{\pm }$ state~\cite{EPAPS,kulic}.

\begin{figure}[t]
\centerline{\includegraphics[width=0.9\columnwidth]{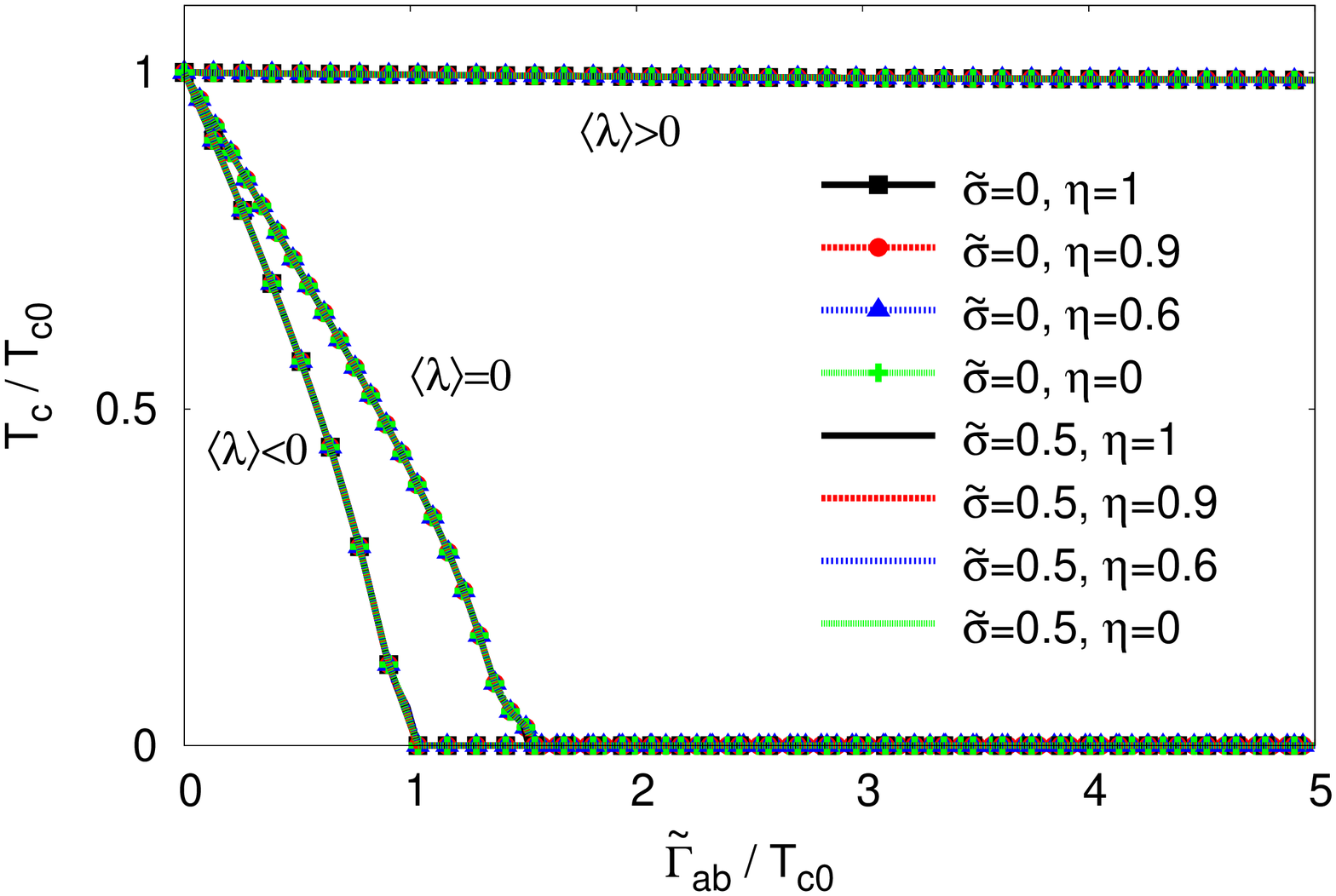}}
\caption{(color online). Critical temperature for various $\tilde{\sigma}$ and $\eta$ as a function of the effective interband scattering rate $\tilde{\Gamma}_{ab}$ for the same parameters. Note that curves for different sets of $\tilde{\sigma}$  and $\eta$ overlap and fall onto one of the three universal curves depending on the $\la \lambda \ra$. $N_{b}/N_{a}=2$, coupling constants for illustrative purpose are chosen  for $\la \lambda \ra>0$ as   $(\lambda_{aa},\lambda_{ab},\lambda_{ba},\lambda_{bb}) = (3,-0.2,-0.1,0.5)$, for $\la \lambda \ra=0$ as $(2,-2,-1,1)$ and for $\la \lambda \ra<0$ as $(1,-2,-1,1)$.
\label{fig1}}
\end{figure}
The linearized Eliashberg equations (\ref{eq:Elias1}) and (\ref{eq:Elias2}) are now solved numerically, varying $T$  and finding $T_{c}$  as the highest temperature where a nontrivial solution appears. Results for $T_{c}$  as a function of $\tilde{\Gamma}_{ab}$, are shown in Fig~\ref{fig1},  in which situation all cases with various values of $\tilde{\sigma}$ and $\eta $ fall on the same universal $T_{c}$ curve for each average $\langle \lambda \rangle \equiv \left[ (\lambda_{aa}+\lambda_{ab})N_{a}/N + (\lambda_{ba}+\lambda_{bb})N_{b}/N\right]$ with $N=N_{a}+N_{b}$. It is clearly seen that depending on the sign of $\langle \lambda \rangle$, one gets two types of $T_{c}$ behavior versus $\tilde{\Gamma}_{ab}$ in the $s_{\pm }$ scenario. For type (i), the critical temperature vanishes at a finite impurity scattering rate $\Gamma_{a}^\mathrm{crit}$ for $\langle \lambda \rangle <0$. For type (ii), $\la \lambda \ra>0$, the critical temperature remains finite at $\Gamma_{a} \rightarrow \infty $. In the marginal case of $\langle \lambda \rangle =0$ we find that $\tilde{\Gamma}^\mathrm{crit}\rightarrow \infty $ but with exponentially small $T_{c}$. Therefore, we have found universal behavior of $T_{c}$ controlled by a single parameter $\langle \lambda \rangle$.
\begin{figure}[tbp]
\centerline{\includegraphics[width=0.9\columnwidth]{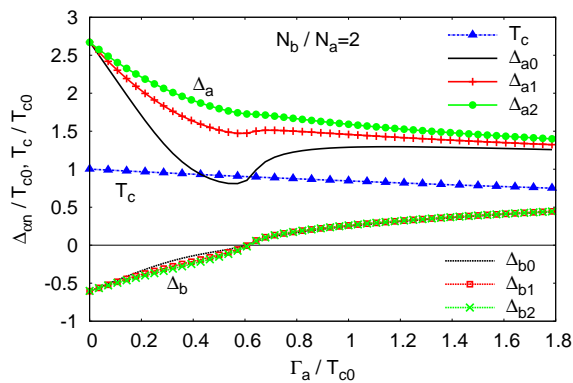}}
\caption{(color online). Critical temperature $T_{c}$ and $\tilde{\Delta}_{\protect\alpha n}$ (in units of $T_{c0}$) for first, second, and third Matsubara frequencies $n$ at $T=0.04T_{c0}$, $\tilde{\protect\sigma}=0.5$, $\protect\eta =1$, and $N_{b}/N_{a}=2$ and $\la \lambda \ra>0 $. The coupling constant are chosen as in Fig.~\ref{fig1}.}
\label{fig2}
\end{figure}

\textit{Weak-coupling limit.} To understand the origin of the two types of
limiting behavior of $T_c$ in an $s_{\pm}$ scenario, we now consider the weak
coupling limit assuming $\lambda_{\alpha \beta}(n -n^{\prime }) =
\lambda_{\alpha \beta} \Theta(\omega_0 -|\omega_n|) \Theta(\omega_0 -
|\omega_n^{\prime }|)$. In this approximation
the calculation can be performed analytically.

We introduce the parameter
\begin{equation}
\Delta_\alpha = \Theta(\omega_0 -|\omega_n|) \sum_{\beta} \lambda_{\alpha
\beta} \pi T \sum_{|\omega_n|<\omega_0} \frac{ \tilde{\Delta}_{\beta n}}{|\omega_{ n}|},
\label{eq:v_g}
\end{equation}
which plays the role of the pair potential in the clean limit. Substituting $%
\tilde\Delta_{\alpha n}$ from Eqs.~(\ref{eq:Elias1}) and (\ref{eq:Elias2}) and recalling $%
\tilde\Gamma_{ab} / \tilde\Gamma_{ba} = N_b/N_a$, we get for $|\omega_n| <
\omega_0$ an equation for $\Delta_\alpha$ similar to AG:
\begin{equation}
\Delta_{\alpha} = \lambda_\alpha {\langle \Delta \rangle} (I_1-I_2) + I_2
\sum_{\beta} \lambda_{\alpha \beta} \Delta_\beta,  \label{eq:2bandAG}
\end{equation}
where $I_1 = \pi T \sum_{|\omega_n|<\omega_0} 1/|\omega_n| \approx \ln {%
2\gamma \omega_0}/{(\pi T_c)}$, $I_2 = \pi T
\sum_{|\omega_n|<\omega_0} 1/(|\omega_n| + \tilde{\Gamma}_{ab} +
\tilde{\Gamma}_{ba})$, $\lambda_\alpha = \sum_\beta
\lambda_{\alpha \beta}$, and ${\langle \Delta \rangle}= (\Delta_a
N_a/N + \Delta_b N_b/N)$.

In the clean limit, $I_2 = I_1$, Eq.~(\ref{eq:2bandAG}) reduces to $%
\Delta_\alpha= I_1 \sum_\beta \lambda_{\alpha \beta} \Delta_\beta$.
Diagonalization of this equation results in the equation for the critical
temperature $I_1 = 1/\lambda_0$, where $\lambda_0 =
(\lambda_{aa}+\lambda_{bb})/2 + \sqrt{(\lambda_{aa}-\lambda_{bb})^2/4 +
\lambda_{ab}\lambda_{ba}} > 0$ is the highest positive eigenvalue of the
matrix $\lambda_{\alpha \beta}$. The critical temperature is then $T_{c0} =
\frac{2\gamma}{\pi} \omega_0 \exp(-1/\lambda_0)$. A similar expression was found in \cite{golubov97,ng}.
The relative sign of the
pair potential of the bands is determined by the off-diagonal interaction
matrix elements: $\mathrm{sgn}\left({\Delta_a/\Delta_b}\right) = \mathrm{sgn}%
\left({\lambda_{ab}}\right)$.

When $\Delta_a - \Delta_b \neq 0$, nonmagnetic impurities suppress the
critical temperature~\cite{golubov97}. The critical value of the impurity
scattering rate for type (i) systems is given by $\ln[\omega_0/(\tilde%
\Gamma_{ab}+\tilde\Gamma_{ba})_\mathrm{crit}] = \langle \lambda \rangle
/(\lambda_{aa} \lambda_{bb} - \lambda_{ab} \lambda_{ba})$.

We now focus on the case of type (ii) systems, $\langle \lambda \rangle > 0$%
, which, to the best of our knowledge, have not been discussed extensively in the literature. Multiplying
both sides of Eq. (\ref{eq:2bandAG}) with $N_{\alpha}$, followed by a
summation and using in the dirty limit $I_2 \to 0$, one obtains $1 = I_1 \langle \lambda \rangle $ and
consequently $T_c = \frac{2\gamma}{\pi} \omega_0 \exp(-1/{\langle \lambda
\rangle})$. Analysis of  Eqs.(\ref{eq:Elias1})  and (\ref{eq:Elias2}) shows that, for small $n$
in the clean limit, $\tilde\Delta_{\alpha n} \sim \Delta_\alpha$, while in
the dirty limit both $\tilde\Delta_{a n}$ and $\tilde\Delta_{b n}$ converge to the same value, $\tilde\Delta_{\alpha n } \to \Delta_{\Gamma \to \infty}$,
that is the $s_{++}$ state is realized. If the initial state corresponds to $s_\pm$, a transition $s_\pm \to s_{++}$ at a finite concentration of impurities must exist.

There is a simple physical argument behind the $s_\pm \to s_{++}$ transition.
With  increasing inter-band disorder, the gap functions on the different Fermi surfaces tend to the same value. A similar effect has been found in
Refs.~\cite{scharnberg,golubov97} for a two-band systems with $s_{++}$ symmetry, and in Ref.~\cite{Mishra} discussing  node lifting on the electron pockets for the extended $s$-wave state in FeSC.

To demonstrate the transition explicitly, we calculate $\tilde{\Delta}%
_{\alpha n}$ for $n=0,1,2$ at $T=0.04T_{c0}$ and show the results in Fig.~%
\ref{fig2} for a particular choice of $\lambda _{\alpha \beta }$ with $%
\langle \lambda \rangle >0$. For this parameter set, $T_{c0} \approx 40$K. Both order parameters $\tilde{\Delta}_{a(b)n}$ converge to $ \Delta_{\Gamma \to \infty}$ for large disorder, while the $T_{c}$ suppression quickly
saturates. The transition $s_{\pm }\rightarrow s_{++}$ provides a
possible explanation for the observed much weaker reduction of the critical
temperature than the naive application of the AG formula.

\begin{figure}[tbp]
\centerline{\includegraphics[width=0.9\linewidth]{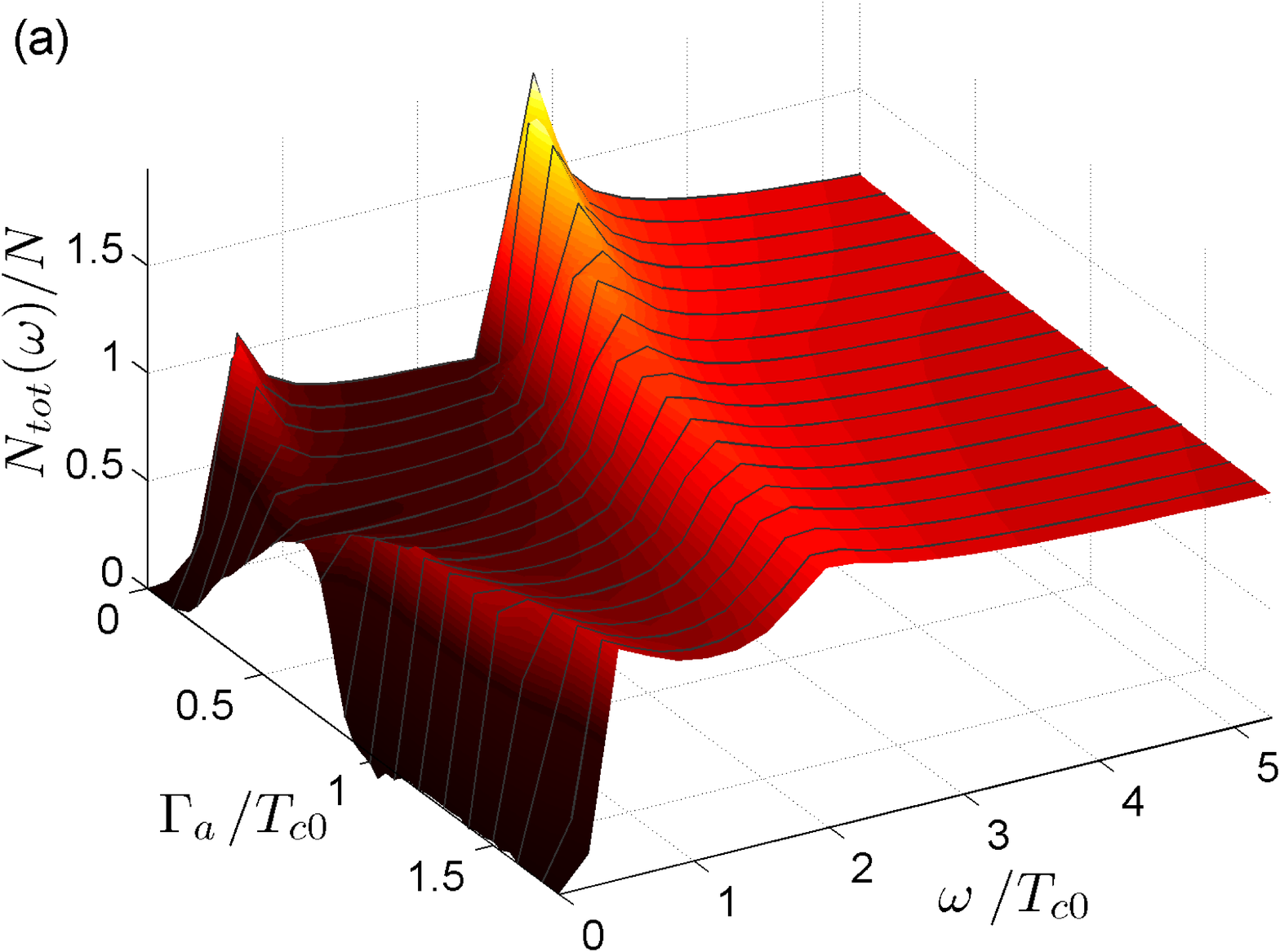}} %
\centerline{\includegraphics[width=0.9\linewidth]{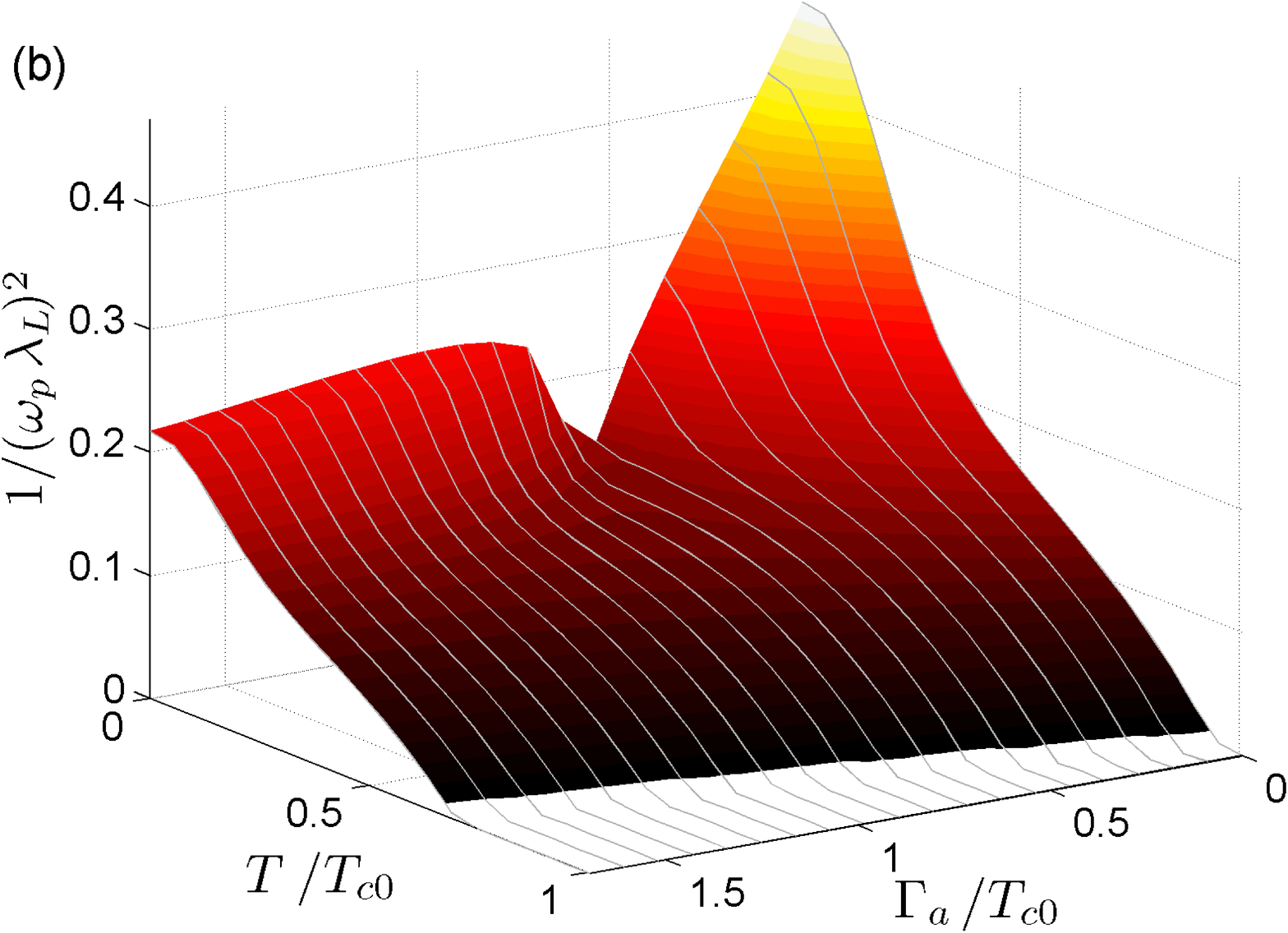}}
\caption{(color online). (a) Density of states $N_{tot}(\protect\omega )/N$
vs. $\Gamma _{a}/T_{c0}$ and $\protect\omega /T_{c0}$ for $\la \lambda \ra>0 $ and impurity parameters $\tilde{\protect\sigma}=0.5$, $\protect\eta %
=1$, $N_{b}/N_{a}=2$, $N=N_{a}+N_{b}$. (b) Total superfluid density $1/(%
\protect\omega _{p}\protect\lambda _{L})^{2}$ vs. $\Gamma _{a}/T_{c0}$ and $%
T/T_{c0}$ where $\protect\omega _{p}$ is the total plasma frequency and $%
\protect\lambda _{L}$ is the London penetration depth.  }
\label{fig3}
\end{figure}

Another important consequence of the transition $s_{\pm }\rightarrow s_{++}$, relevant to experiments in pnictides, is gapless
superconductivity as one of the gaps vanishes. The density of states $%
N_{tot}(\omega )=-\sum_{\alpha }\mathrm{Im}g_{0\alpha }(\omega )/\pi $ is
shown in Fig.~\ref{fig3}(a) for a type (ii) case. With increasing impurity
scattering rate, the lower gap is seen to close, leading to a finite
residual $N_{tot}(\omega =0)$, followed by a reopening of the gap. A similar
behavior is reflected in the temperature dependence of the penetration
depth, [Fig.~\ref{fig3}(b)], which varies in the clean limit with activated
behavior controlled by the smaller gap, crossing over to $T^{2}$ in the
gapless regime, to a new activated behavior in the $s_{++}$ state in the
dirty limit. Figure~\ref{fig3}(b) should be compared to similar works, where
the effect of scattering on the $T$-dependent superfluid density was
calculated for a two-band $s_{++}$ state~\cite{golubov02,mazin,scharnberg}.

\textit{Conclusions.} We have shown that in two-band models with an $s_\pm$
ground state, $T_c$ has a universal dependence on the impurity scattering
rate which can be calculated explicitly in terms of the interband to intraband
impurity scattering rate ratio. We demonstrated that $s_\pm$
superconductivity may be quite robust against nonmagnetic impurities,
depending on the ratio of interband to intraband pairing coupling constants,
and may even display a transition to an $s_{++}$ gap structure with
increasing disorder, which will manifest itself in thermodynamic and
transport properties.

 The authors are grateful to S.-L. Drechsler, P.
Fulde, I.I. Mazin, V. Mishra, and D.J. Scalapino for useful discussions. The
present work was partially supported by the DFG Priority Programme SPP1458
(D.V.E), Dutch FOM (A.A.G), DOE DE-FG02-05ER46236 (P.J.H and M.M.K), and RFBR
09-02-00127, Presidium of RAS program N5.7, FCP GK P891, and GK
16.740.12.0731, and President of Russia MK-1683.2010.2 (M.M.K). P.J.H and M.M.K are grateful for the support of the Kavli Institute for Theoretical Physics and the Stanford Institute for Materials \& Energy Science during the writing of this work.

\cleardoublepage
\onecolumngrid

\setcounter{figure}{0}
\setcounter{equation}{0}
\renewcommand\theequation{S\arabic{equation}}
\renewcommand\thefigure{S\arabic{figure}}
\newcommand{\etal}{\textit{et al.}}
\newcommand{\BKFA}{Ba$_{0.68}$K$_{0.32}$Fe$_2$As$_2$\xspace}
\newcommand{\BFMA}{$\rm Ba(Fe_{0.88}Mn_{0.12})_2As_2$}
\newcommand{\BFA}{BaFe$_2$As$_2$\xspace}
\newcommand{\KFA}{KFe$_2$As$_2$\xspace}
\newcommand{\mJ}{mJ/molK$^2$\xspace}

\section{Supplementary online material for the article ``Disorder induced
transition between $s_{\pm }$ and $s_{++}$ states in two-band superconductors''}

%\newpage \widetext

\textit{Multiband system in the weak coupling approximation.}

The Eliashberg equations  in the general form on the imaginary Matsubara axis are
\[
\tilde{\phi}_{an}=\pi T\sum_{n^{\prime }}\sum_{i}\lambda
_{ai}(n-n^{\prime })\frac{\tilde{\phi}_{in^{\prime }}}{\sqrt{\tilde{%
\omega}_{in^{\prime }}^{2}+\tilde{\phi}_{in^{\prime }}^{2}}}+\sum_{i}\tilde{%
\Gamma}_{ai}\frac{\tilde{\phi}_{in}}{\sqrt{\tilde{\omega}_{in}^{2}+\tilde{%
\phi}_{in}^{2}}}
\]%
\[
\tilde{\omega}_{an}=\omega +\pi T\sum_{n}\sum_{i}|\lambda _{ai}(n-n^{\prime })|%
\frac{\tilde{\omega}_{in^{\prime }}}{\sqrt{\tilde{\omega}_{in^{\prime
}}^{2}+\phi _{in^{\prime }}^{2}}}+\sum_{i}\tilde{\Gamma}_{ai}\frac{\tilde{%
\omega}_{in}}{\sqrt{\tilde{\omega}_{in}^{2}+\tilde{\phi}_{in}^{2}}}.
\]%
We consider the weak coupling limit (the so-called $\Theta \Theta$-model), which
corresponds to: $\lambda _{\alpha \beta }(n-n^{\prime })=\lambda _{\alpha
\beta }\Theta (\omega _{0}-|\omega _{n}|)\Theta (\omega _{0}-|\omega
_{n}^{\prime }|)$. In this case, the term with $\lambda _{ai}(n-n^{\prime })$ vanishes,
\[
\tilde{\phi}_{an}=\Theta (\omega _{0}-|\omega _{n}|)\sum_{i}\lambda
_{ai}\pi T\sum_{\left\vert \omega _{n^{\prime }}\right\vert \leq \omega
_{0}}\frac{\tilde{\phi}_{in^{\prime }}}{\sqrt{\tilde{\omega}_{in^{\prime
}}^{2}+\tilde{\phi}_{in^{\prime }}^{2}}}+\sum_{i}\tilde{\Gamma}_{ai}\frac{%
\tilde{\phi}_{in}}{\sqrt{\tilde{\omega}_{in}^{2}+\tilde{\phi}_{in}^{2}}}
\]%
\[
\tilde{\omega}_{an}=\omega _{n}+\sum_{i}\tilde{\Gamma}_{ai}\frac{\tilde{%
\omega}_{in}}{\sqrt{\tilde{\omega}_{in}^{2}+\tilde{\phi}_{in}^{2}}}.
\]%
or introducing
\[
Z_{an}=1+\sum_{i}\frac{\tilde{\Gamma}_{ai}}{\sqrt{\omega _{n}^{2}+\tilde{%
\Delta}_{in}^{2}}}.
\]%
\begin{equation}
\tilde{\Delta}_{an}=\Theta (\omega _{0}-|\omega _{n}|)\sum_{i}\lambda
_{ai}\pi T\sum_{\left\vert \omega _{n^{\prime }}\right\vert \leq \omega
_{0}}\frac{\tilde{\Delta}_{in^{\prime }}}{\sqrt{\omega _{n^{\prime }}^{2}+%
\tilde{\Delta}_{in^{\prime }}^{2}}}+\sum_{i}\tilde{\Gamma}_{ai}\frac{\tilde{%
\Delta}_{in}-\tilde{\Delta}_{an}}{\sqrt{\omega _{n}^{2}+\tilde{\Delta}%
_{in}^{2}}}.
%\tag{S1}
\label{Swg}
\end{equation}

Note that intraband nonmagnetic impurities scattering rate $\tilde{\Gamma}_{aa}$ drop out from this equation according to Anderson's theorem.
Let us consider for simplicity $T=T_{c}$. We would also like to split the gap functions into two parts: part $A$ undergoes a strong impurity scattering between  different parts of the Fermi surface separated by
large wave vectors (say, electron and hole bands) with $\tilde\Gamma_{ij} \propto u_1N_j$ ,
while the part $B$ undergoes weak scattering with $\tilde\gamma_{ij} \propto u_2 N_{j}$. In this case, Eqs.~(\ref{Swg}) for $a\in A$ become
\begin{eqnarray}
\tilde{\Delta}_{an} &=&\Theta (\omega _{0}-|\omega _{n}|)\sum_{i}\lambda
_{ai}\pi T_{c}\sum_{\left\vert \omega _{n^{\prime }}\right\vert \leq \omega
_{0}}\frac{\tilde{\Delta}_{in^{\prime }}}{|\omega _{n^{\prime }}|}  \nonumber
\\
&&+\sum_{j\in A}\tilde{\Gamma}_{aj}\frac{\tilde{\Delta}_{jn}-\tilde{\Delta}%
_{an}}{|\omega _{n}|}+\sum_{l\in B}\tilde{\gamma}_{al}\frac{\tilde{\Delta}%
_{ln}-\tilde{\Delta}_{an}}{|\omega _{n}|},
\end{eqnarray}
For $|\omega _{n}|\leq \omega_{0}$, the solution of this equation can be written as
\begin{equation}
\tilde{\Delta}_{an}=\frac{\sum_{j\in A}\tilde{\Gamma}_{aj}\tilde{\Delta}_{jn}+\sum_{l\in
B}\tilde{\gamma}_{al}\tilde{\Delta}_{jn}+I_{a}|\omega _{n}|}{|\omega _{n}|+\sum_{j\in A} \tilde{\Gamma}_{aj}+\sum_{l\in
B}\tilde{\gamma}_{al}},  \label{main}
\end{equation}
where
$$
I_{a} =\pi T_{c} \sum_i\lambda _{ai} \sum_{|n^{\prime }|\leq N_{1}}
\frac{\tilde{\Delta}_{in^{\prime }}}{|\omega _{n^{\prime }}|} , \,\,\,\,\,\, (N_{1}\simeq \omega _{0}/2\pi T_{c0}\gg 1).
$$

Since $\tilde{\Gamma}_{aj} \propto u_{1}N_{j}$, for $u_{1} \rightarrow \infty$ from Eq.~(\ref{main}) we see that the gap function coincides with $\tilde{\Delta}_{an} \propto \sum_{j\in 1} N_{j} I_{j}$.
 On the other hand, the gap functions belonging to the other part (weakly coupled bands) can have different signs.

\textit{$T_c$ dependence on the effective interband scattering rate.}

One finds that $\tilde{\Gamma}_{ab(ba)} = \Gamma _{a(b)} f(\tilde\sigma,\eta)$

with
\begin{equation}
f(\tilde\sigma,\eta) = \frac{(1-\tilde{\sigma})}{\tilde{\sigma}(1-\tilde{\sigma})\eta \frac{(N_{a}+N_{b})^{2}}{N_{a}N_{b}}+(\tilde{\sigma}\eta -1)^{2}},
\label{eq:f}
\end{equation}
where $\tilde{\sigma}=(\pi ^{2}N_{a}N_{b}u^{2})/(1+\pi ^{2}N_{a}N_{b}u^{2})$
and $\Gamma _{a(b)}=
%2
n_{imp}\pi N_{b(a)}u^{2}(1-\tilde{\sigma})$ are generalized cross-section
and normal state scattering rate parameters, respectively, and $n_{imp}$ is the impurity concentration. The parameter $\eta$ controls the ratio of intraband and interband scattering as $v=u\eta$. In the Born  limit, $\tilde{\sigma}\rightarrow 0$, while for $\tilde{\sigma}\rightarrow 1$ the unitary limit  is achieved. Function $f(\tilde\sigma,\eta)$ is limited from above by 1, therefore, in the unitary limit nonmagnetic impurities do not affect $T_{c}$.
\begin{figure}[t]
\centerline{\includegraphics[width=0.7\columnwidth]{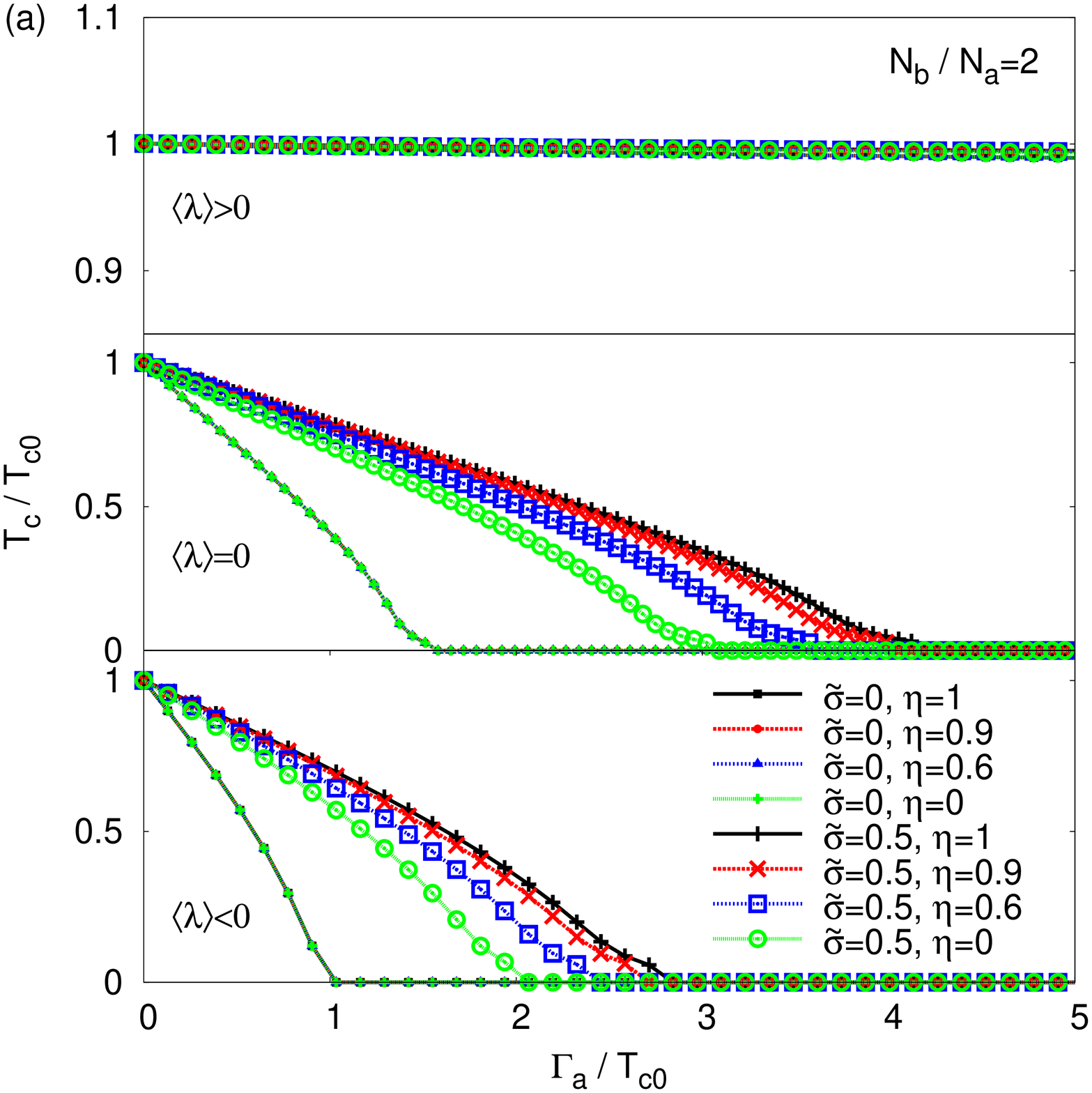}}
\centerline{\includegraphics[width=0.7\columnwidth]{TcGammaUni_new.eps}}
\caption{(color online). Critical temperature for various $\tilde{\sigma}$ and $\eta $ as a function of (a) the impurity scattering rate $\Gamma_{a}$ and (b) the effective interband scattering rate $\tilde{\Gamma}_{ab}$ for the same parameters. $N_b/N_a=2$, coupling constants are the same as in Fig.1 of the main text.
\label{fig1}}
\end{figure}
From Eq.~(\ref{eq:f}), we therefore recover explicitly the well-known but counterintuitive result that in the unitary limit nonmagnetic impurities do not affect $T_{c}$ in the $s_{\pm}$ state.

Results for $T_{c}$ as a function of the pairbreaking parameter $\Gamma _{a}$ (proportional to impurity concentration $n_{imp}$) are shown in Fig.~\ref{fig1}(a). For illustrative purposes, the coupling constants $\lambda_{\alpha \beta}$ are chosen the same as in Fig.~1 of the main text. Note that the strongest suppression is generally found for pure uniform scattering, $\eta=0$, in the Born limit ($\tilde{\sigma}\rightarrow 0$)

and that in the opposite limit of pure intraband scattering, $u=0$, we have $\eta \rightarrow \infty $, so that there is no pairbreaking since $\tilde{\Gamma}_{ab} \rightarrow 0$. The similar situation takes place for the strong unitary limit.
It is clearly seen that depending on the sign of the average $\langle \lambda \rangle \equiv \left[ (\lambda_{aa}+\lambda_{ab})N_{a}/N+(\lambda _{ba}+\lambda _{bb})N_{b}/N\right]$ with $N=N_{a}+N_{b}$, one gets two types of $T_{c}$ vs. $\Gamma_a$ behavior in the $s_{\pm}$ scenario. For type (i), the critical temperature vanishes at a finite impurity scattering rate $\Gamma_{a}^\mathrm{crit}$ for $\langle \lambda \rangle < 0$. For type (ii), the critical temperature remains finite at $\Gamma_{a} \rightarrow \infty$. In the marginal case of $\langle \lambda \rangle =0 $ we find that $\Gamma_a^\mathrm{crit} \rightarrow \infty $, but with
exponentially small $T_{c}$. Fig.~\ref{fig1}(b) is equivalent to Fig.~1 of the main text and shown here for easier comparison with the panel (a).

\end{document}